\documentclass[10pt,a4paper]{book}
\usepackage{frontier07}
\newcommand{\be}{\begin{eqnarray}}
\newcommand{\ee}{\end{eqnarray}}

\begin{document}
\newcounter{ctr}
\setcounter{ctr}{\thepage}
\addtocounter{ctr}{8}

\talktitle{Primordial antimatter in the contemporary universe}
\talkauthors{Cosimo Bambi}

\begin{center}
{Dipartimento di Fisica, Universit\`a degli Studi di Ferrara \\
via Saragat 1, 44100 Ferrara, Italy}
\end{center}

\shorttitle{Search for primordial antimatter} 

\firstauthor{C. Bambi}

\begin{abstract}
In some baryogenesis scenarios, the universe acquires a 
non-vanishing average baryonic charge, but the baryon to 
photon ratio is not spatially constant and can be even 
negative in some space regions. This allows for existence 
of lumps of antimatter in our neighborhood and the 
possibility that very compact antimatter objects make a 
part of cosmological dark matter. Here I discuss the 
peculiar signatures which may be observed in a near future.
\end{abstract}

One can conclude from simple considerations that there is 
much more matter than antimatter around us~\cite{steigman}.
However, the origin of matter--antimatter asymmetry in the
universe is unknown: the Standard Model of particle physics 
is certainly unable to explain it and new physics is 
necessary~\cite{ad-cp}. Assuming a homogeneous and
isotropic universe, from the Big Bang Nucleosynthesis 
(BBN)~\cite{pdg} and the Cosmic Microwave Background 
Radiation (CMBR)~\cite{wmap} one can determine the baryon 
to photon ratio $\beta$ 
\be
\beta = \frac{n_B-n_{\bar B}}{n_\gamma} 
\approx 6 \cdot 10^{-10} 
\label{eta-obs}
\ee
where $n_B \gg n_{\bar B}$. On the other hand, the freeze-out 
abundances in a homogeneous baryo-symmetric universe would be
$n_B/n_\gamma = n_{\bar B}/n_\gamma \sim 10^{-18}$~\cite{zeld65}.

However, Eq.~(\ref{eta-obs}) may not be the end
of the story. One can indeed distinguish three main types
of cosmological matter--antimatter asymmetry:
\begin{enumerate}
\item {\bf Homogeneous matter dominated universe}. Here $\beta$ 
is constant and the universe is 100\% matter dominated. This 
is certainly the most studied case (see e.g. Refs.~\cite{review, freese}) 
but it is not very interesting for astrophysical observations, 
because there is only one observable quantity, $\beta$, which 
cannot contain much information on high energy physics. 
\item {\bf Globally B-symmetric universe}. Such a possibility 
appears quite reasonable and ``democratic'': the universe would 
consist of equal amount of similar domains of matter and 
antimatter. However, it seems observationally excluded or, to 
be more precise, the size of the domain where we live should be 
at least comparable to the present day cosmological horizon~\cite{cdg}.
So, even in this case observations cannot determine nothing but $\beta$.
\item {\bf Inhomogeneous matter dominated universe}. In this
case the universe has a non-vanishing baryonic charge, but
$\beta$ is not spatially constant and can even be negative in
some space regions. Lumps of antimatter can be scattered 
throughout the universe.
\end{enumerate}

Here I will discuss possible observational signatures of the
third case: even if at first glance such a picture may appear 
strange, just because we are used to think about ordinary matter
around us, there are no theoretical and experimental reasons to
reject it. At present, the source of CP violation responsible 
for the observed B-asymmetry in the universe is unknown, so
generation of lumps of antimatter is not so exotic as one may 
naively think. Moreover, compact antimatter objects can easily 
survive in a matter dominated universe up to the present days. 
The talk is based on a work made in collaboration with Alexander 
Dolgov~\cite{noi}. The reference baryogenesis mechanism is the
one in~\cite{ad-silk}. The phenomenology of other scenarios
can be found in Refs.~\cite{khlop, zhit}.

\section{Baryogenesis framework}

Let us now briefly review the baryogenesis framework 
suggested in Ref.~\cite{ad-silk}. The basic ingredient is
the Affleck-Dine mechanism~\cite{a-d}, where a scalar field 
$\chi$ with non-zero baryonic charges have the potential with 
flat directions, that is directions along which the 
potential energy does not change. Due to the infrared instability 
of light fields in de Sitter spacetime~\cite{infrared},
during inflation $\chi$ can condense along the flat directions
of the potential, acquiring a large expectation value. 
In the course of the cosmological expansion, the Hubble
parameter drops down and, when the mass of the field exceeds 
the universe expansion rate, $\chi$ evolves to the equilibrium 
point and the baryonic charge stored in the condensate 
is transformed into quarks by $B$-conserving 
processes. Since here CP is violated stochastically by a 
chaotic phase of the field $\chi$, then during the motion 
to the equilibrium state the matter and antimatter domains 
in a globally symmetric universe would be created. An 
interesting feature of the model is that regions with a very 
high $\beta$, even close to one, could be formed.

If the scalar field $\chi$ is coupled to the inflaton $\Phi$
with an interaction term of the kind
$V(\chi,\Phi) = \lambda |\chi|^2 (\Phi - \Phi_1)^2$,
the ``gates'' to the flat directions might be open only for
a short time when the inflaton field $\Phi$ was close to 
$\Phi_1$. In this case, the probability of the penetration 
to the flat directions is small and $\chi$ could acquire a 
large expectation value only in a tiny fraction of space. 
The universe would have a homogeneous background of baryon 
asymmetry $\beta \sim 6 \cdot 10^{-10}$ generated by the same 
field $\chi$, which did not penetrate to larger distance 
through the narrow gate, or by another mechanism of baryogenesis, 
while the high density matter, $\beta >0$, and antimatter, 
$\beta < 0$, regions would be rare, although their contribution to 
the cosmological mass density might be significant or even 
dominant. In the simple model of Ref.~\cite{ad-silk}, such
high density bubbles could form clouds of matter or antimatter 
and more compact object like stars, anti-stars or primordial 
black holes. In the non-collapsed regions, primordial 
nucleosynthesis proceeded with large $|\beta|$, producing nuclei 
heavier than those formed in the standard BBN~\cite{ibbn}.

\section{Phenomenology \label{s-late}}

In what follows I will not dwell on possible scenarios
of antimatter creation, but simply consider phenomenological
consequences of their existence in the present day universe,
in particular in the Galaxy. Some considerations on the
cosmological evolution of lumps of antimatter in a baryon
dominated universe can be found in Refs.~\cite{noi,zhit}.

\subsection{Indirect detection}

The presence of anti-objects in the Galaxy today should lead 
to the production of the gamma radiation from matter--antimatter 
annihilation. Hence we would expect $\sim 100$ MeV $\gamma$ 
from the decay of $\pi^0$ mesons produced in $p \bar p$ 
annihilation, with an average of 4 $\gamma$ per annihilation, 
and  $2\,\gamma$ from $e^+e^-$ annihilation with $E=0.511 $ MeV, 
if $e^+ e^-$ annihilate at rest. In addition to the slow 
background positrons, there should be also energetic secondary 
positrons produced by pion decays from $p \bar p$ annihilation. 
Astronomical observations are seemingly more sensitive to 
$p \bar p$ annihilation because the total energy release in 
$p \bar p$ annihilation is 3 orders of magnitude larger than 
that in $e^+e^-$ annihilation and the galactic gamma ray 
background at 100 MeV is several orders of magnitude lower than 
the one at 0.5 MeV. On the other hand, $e^+e^-$ annihilation 
gives the well defined line which is easy to identify.

For compact anti-objects like anti-stars, one find that the
size of the anti-object, $R$, is much larger than the proton 
or electron mean free path inside the anti-object, 
$\lambda_{free} \sim 1/(\sigma_{ann}\, n_{\bar p})$, where
$\sigma_{ann}$ is the annihilation cross section for $p \bar p$ 
or $e^+e^-$ (they have similar order of magnitude) and 
$n_{\bar p}$ is the antiproton number density in the anti-object.  
In this case, the annihilation takes place on the surface,
all the protons and electrons that hit the surface of the 
anti-object annihilate and the annihilation cross section is 
given by the geometrical area of the anti-object, that is
$\sigma = 4 \pi R^2$. The gamma ray luminosity of such a 
compact anti-object is
\be
L_\gamma \approx 10^{27} \, 
\left( \frac{R}{R_\odot} \right)^2
\left( \frac{n_p}{{\rm cm}^{-3}} \right) 
\left(\frac{v}{10^{-3}}\right) \, {\rm erg/s} \, ,
\label{L-sur}
\ee
where $R_\odot \sim 7 \cdot 10^{10}$ cm is the Solar radius
and $n_p v$ is the proton flux. With this luminosity, a solar mass 
anti-star would have the life time of the order of $10^{27}$ s 
(considering only matter--antimatter annihilation), if all the 
factors in Eq.~(\ref{L-sur}) are of order unity. For an anti-star 
in the galactic disc, the $\gamma$ flux observable on the Earth 
would be
\be\label{flux-anti-star}
\phi_{Earth} \sim 10^{-7} \,
\Big(\frac{R}{R_\odot}\Big)^2
\Big(\frac{1 \, \rm pc}{d}\Big)^2 \; 
{\rm cm^{-2} \; s^{-1}} \, .
\ee
where $d$ is the distance of the anti-star from the Earth.
Such a flux should be compared with the point source 
sensitivity of EGRET~\cite{egret}, at the level of 10$^{-7}$ 
photons cm$^{-2}$ s$^{-1}$ for $E_\gamma > 100$ MeV, and of 
the near-future GLAST~\cite{glast}, which should be about two 
order of magnitude better, i.e. $\sim 10^{-9}$ photons 
cm$^{-2}$ s$^{-1}$. So, anti-stars should be quite close to 
us in order to be detectable point-like sources and their
observation would result difficult if they were very compact 
objects, as e.g. anti-neutron stars. On the other hand, if 
such an anti-star lived in the galactic center, where 
$n_p \gg 1/$cm$^3$, its luminosity would be larger. 
Anomalously bright lines of 0.5 MeV are observed recently in the 
galactic center~\cite{gr-an}, galactic bulge~\cite{e-continuum} 
and possibly even in the halo~\cite{e-line}. Though an excess 
of slow positrons is explained in a conventional way as a 
result of their creation by light dark matter particles, such a 
suggestion is rather unnatural, because it requires a fine-tuning 
of the mass of the dark matter particle and the electron mass. 
More natural explanation is the origin of these positrons from 
primordial antimatter objects.

The existence of primordial antimatter in the Galaxy would 
increase the galactic diffuse gamma ray background as well. 
Standard theoretical predictions and observational data agree 
on a galactic production rate of $\gamma$ in the energy 
range $E_\gamma > 100$ MeV~\cite{noi}
\be\label{total}
\Gamma_\gamma^{tot} \sim 10^{43} \; {\rm s^{-1}} \; .
\ee
Requiring that annihilation processes on anti-stars surface
cannot produce more than 10\% of the standard galactic production 
rate~(\ref{total}), we obtain the following bound on the 
present number of anti-stars
\be
N_{\bar S} \lesssim 10^{12} \Big(\frac{R_\odot}{R}\Big)^2 \; ,
\label{bound-gamma}
\ee
where, for simplicity, we assumed that all the anti-stars 
have the same radius $R$. However the constraint is not very
strong: for solar type anti-stars, their number cannot exceed 
the one of ordinary stars!

Let us now consider the annihilation of antimatter from the 
anti-stellar wind with protons in the interstellar medium. 
Since the number of antiprotons reached a stationary value, 
the production rate of 100 MeV $\gamma$ in the Galaxy has to 
be proportional to $N_{\bar S}$. The luminosity of the Galaxy 
in 100 MeV $\gamma$ rays from anti-stellar wind would be 
$L_{\bar S} \sim 10^{44} W\,N_{\bar S}/N_S$ erg/s, where
$W$ is the anti-stellar wind to solar wind flux ratio. Since 
from Eq.~(\ref{total}) we find that the total Galaxy luminosity 
in 100 MeV $\gamma$ is $L_{\gamma}^{tot} \sim 10^{39}$ erg/s, 
the related bound on the anti-star to star number ratio is
$N_{\bar S}/N_S \lesssim 10^{-6} \, W^{-1}$, always assuming 
that the contribution from new physics cannot exceed 10\% of 
$L_{\gamma}^{tot}$. A similar restriction can also 
be obtained from the 0.511 MeV line created by $e^+ e^-$
annihilation with positrons from the anti-stellar wind.

On the other hand, if anti-stars were formed in the 
very early universe in the regions with a high antimatter 
density~\cite{ad-silk}, such primordial stars would most 
probably be compact ones, like white dwarfs or neutron 
stars. The stellar wind in this case would be much 
smaller that the solar one, $W\ll 1$. Their luminosity from 
the annihilation on the surface should be very low, because 
of their small radius $R$, and their number in the Galaxy may 
be even larger than the number of the usual stars. This 
possibility is not excluded by the previous bounds. Such 
compact dark stars could make a noticeable part of the 
cosmological dark matter.

\subsection{Direct detection}

It is common belief that the abundances of most elements in 
the cosmic rays reflect relative abundances in the Galaxy.
Hence, as the simplest working hypothesis we can assume that
the antimatter--matter ratio in cosmic rays is more or less equal 
to the anti-star--star ratio $N_{\bar S}/N_S$, at least if the 
anti-stars are of the same kind as the stars in the Galaxy.

As for antiprotons and positrons, they cannot be direct 
indicators for the existence of primordial antimatter, because 
they can be produced in many astrophysical processes. For example, 
the observed ${\bar p}/p$ ratio is at the level of 10$^{-4}$ 
and is compatible with theoretical predictions for $\bar p$ 
production by the high energy cosmic ray collisions with the 
interstellar medium. A possible contribution of $\bar p$ from 
primordial lumps of antimatter is not more than about 10\% of 
the total observed $\bar p$ flux, so $N_{\bar S}/N_S \lesssim 10^{-5}$
and the number of anti-stars $N_{\bar S}$ has to be no more
than $10^6$, since the number of ordinary stars in the Galaxy is 
$N_S \sim 10^{11}$.

On the other hand, the possibility of producing heavier 
anti-nuclei (such as anti-helium) in cosmic ray collisions 
is completely negligible and a possible future detection 
of the latter would be a clear signature of antimatter 
objects. At present there exists an upper limit on the 
anti-helium to helium ratio in cosmic rays, at the level 
of 10$^{-6}$~\cite{bess}, leading to the constraint 
$N_{\bar S} \lesssim 10^5$. Such an upper limit can
probably be lowered by 2 or 3 orders of magnitude in a
near future, thanks to AMS~\cite{ams} and 
PAMELA~\cite{pamela} space missions. I would like to 
stress that here we are not assuming that these possible
anti-helium nuclei were produced by nuclear fusion inside 
anti-stars, but that original anti-helium abundance inside 
anti-stars is roughly equal to the helium abundance inside 
ordinary stars. This is certainly a conservative picture,
since anti-stars were formed in high density regions of
the early universe, where the primordial nucleosynthesis produced
much more anti-helium and heavier anti-nuclei~\cite{ibbn}. On 
the other hand, if anti-stars were compact ones from the very 
beginning, the stellar wind from them and the shortage of 
anti-supernova events would spread much less anti-helium 
than the normal stars.

\subsection{More exotic events}

The presence of anti-stars in the Galaxy could lead to
extraordinary events of star--anti-star annihilation. As 
a matter of fact, the radiation pressure produced in the 
collision prevents their total destruction. Still the 
released energy can be huge.

The most spectacular phenomenon is a collision between a 
star and an anti-star with similar masses $M$. A simple
estimate of the amount of the annihilated matter in such 
a collision is $m_{ann} \sim Mv^2$~\cite{noi},
where $v$ is the typical value of the relative velocity and
is about $10^{-3}$. The total energy release would be
$E \sim 10^{48} \, {\rm erg} (M/M_\odot) (v/10^{-3})^2$. 
Most probably the radiation would be emitted in a narrow disk 
along the boundary of the colliding stars. The collision time 
is $t_{coll} \sim R $ and for the solar type star this time is 
about 3 s. The energy of the radiation should be noticeably 
smaller than 100 MeV, because the radiation should degrade in 
the process of forcing the star bounce. This makes this 
collision similar to gamma bursts, but unfortunately some other 
features do not fit so well: the released energy should be 
much larger, about $10^{53}\,\sqrt{v}$~erg and it is difficult 
to explain the features of the afterglow.

\section{Conclusion \label{s-conclusion}}

Unfortunately there are no true conclusions because we are
unable to make clear predictions. However this is the problem of 
all the baryogenesis models: the physics responsible for the
matter--antimatter asymmetry in the universe is unknown and
common approaches are based on the construction and investigation
of toy-models which contain free parameters that
we can only partially constrain with 
the observed asymmetry~(\ref{eta-obs}). Moreover,
most baryogenesis scenarios are based on physics at very
high energy, which will be hardly tested in a near future
by man-made colliders. On the other hand, if we are lucky 
and able to get evidences of the existence of primordial 
antimatter object, the latter will tell us much interesting 
information on high energy physics (CP violation,
B violation, etc.) and, maybe, even on cosmological open
questions such as the nature of dark matter.

Gamma rays from $p \bar p$ annihilation may be observable with 
future or even with existing $\gamma$-telescopes. Quite promising 
for discovery of cosmic antimatter are point-like sources of gamma 
radiation; the problem is to identify a source which is 
suspicious to consist of antimatter. The 100 MeV gamma ray 
background does not have pronounced features which would 
unambiguously tell that the photons came from the annihilation 
of antimatter. The photons produced as a result of $p \bar p$
annihilation would have a well known spectrum but it may be
difficult to establish a small variation of the conventional 
spectrum due to such photons. In contrast, the 0.511 MeV line must 
originate from $e^+e^-$ annihilation and it is tempting to conclude 
that the observed excessive signal from the Galaxy and, especially, 
from the galactic bulge comes from astronomical antimatter objects.
If an anti-star happens to be in the galactic center, its luminosity
from the surface annihilation of the background matter should be
strongly enhanced due to the much larger density of the interstellar 
matter there. So the search of the antimatter signatures in
the direction of the center is quite promising. There is also a 
non-negligible chance to detect cosmic anti-nuclei and not only 
light anti-helium but also much heavier ones, especially if 
anti-stars became early supernovae.

\end{document}